\documentclass[twocolumn,twoside,slac_two]{revtex4}
\usepackage{graphicx}
\usepackage{fancyhdr}
\begin{document}

\title{Hadronic \boldmath{$B$} Decays in PQCD}

%

\author{Satoshi Mishima}
\affiliation{School of Natural Sciences, Institute for Advanced Study,
Princeton, NJ 08540, U.S.A.}

\begin{abstract}
I review recent progress on exclusive hadronic $B$ meson decays in the
 perturbative QCD approach, with focus on puzzles in the branching
 ratios and the CP asymmetries of the $B\to\pi K$ and $B\to\pi\pi$
 modes, and polarization fractions in $B\to VV$ modes.  
\end{abstract}

\maketitle

\thispagestyle{fancy}


\section{Introduction}

$B$ factory experiments have accumulated a lot of data and have reported
many interesting results~\cite{HFAG}.  Some observables, mixing-induced
CP asymmetries for $b\to s$ penguin modes, branching ratios and direct
CP asymmetries for $B\to \pi K$ and $\pi\pi$, and polarization fractions
for penguin-dominated $B\to VV$ modes, have exhibited some deviations
from na\"{\i}ve expectations in the Standard Model.

It is necessary to go beyond na\"{\i}ve estimations for understanding the
observed deviations.  The perturbative QCD (PQCD)
approach~\cite{KLS,Lu:2000em} is one of the theoretical attempts
to include subdominant contributions, such as spectator and annihilation
diagrams, and higher-order corrections.  PQCD has applied to various
two-body $B$ decays at leading order (LO) in $\alpha_s$ and has made
reasonable predictions for various decay modes. Recently, the important
next-to-leading-order (NLO) contributions were evaluated in the 
$B\to\pi K$, $\pi\pi$, and $ \rho\rho$ decays to investigate the
discrepancies between the LO PQCD predictions and the
data~\cite{Li:2005kt,Li:2006cv}.

This talk is organized as follows:  In Sec.~\ref{sec:pqcd}, I briefly
review the PQCD factorization formula.  I discuss the branching ratios
and the CP asymmetries of the $B\to\pi K$ and $\pi\pi$ decays with the
NLO corrections in Sec.~\ref{sec:pik}.  The LO PQCD predictions of the
polarization fractions are presented in Sec.~\ref{sec:vv}.
Section~\ref{sec:summary} is a summary.

\section{PQCD Factorization Theorem\label{sec:pqcd}} 

Most of the calculations of $B$ decay amplitudes rely on the
factorization of decay amplitudes into a product of short-distance and
long-distance physics.  QCD-improved factorization (QCDF)~\cite{BBNS}
and soft-collinear effective theory (SCET)~\cite{SCET} are based on
collinear factorization theorem, but PQCD is based on $k_T$
factorization theorem.

Employing collinear factorization theorem, some decay amplitudes involve
a singularity arising from the end-point region of parton momentum
fractions.  An end-point singularity implies that a decay amplitude is
dominated by soft dynamics and cannot be factorized.  Such soft
contributions are regarded as phenomenological parameters, which can be
fitted from the experimental data.

\begin{figure}[t]
\centering
\includegraphics{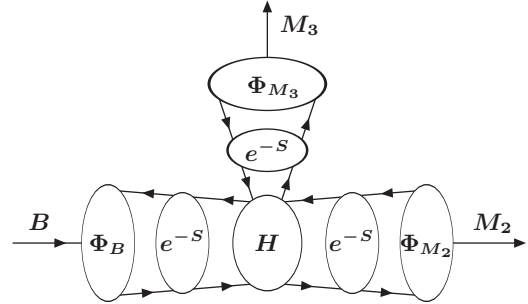}
\caption{Factorization of hadronic two-body $B$ meson decays in the PQCD
 approach.  
\label{fig:fact}} 
\end{figure}

In the PQCD approach with $k_T$ factorization theorem, the Sudakov
factor ensures the absence of the end-point
singularities~\cite{Li:1992nu}.  All amplitudes can be factorizable into
parton distribution amplitudes $\Phi$, the Sudakov factors $e^{-S}$, and
a hard kernel $H$:
\begin{eqnarray}
{\cal A}(B\to M_2 M_3)
\, =\, 
\Phi_{M_2} \otimes \Phi_{M_3} \otimes 
H \otimes e^{-S} \otimes 
\Phi_B 
\;,
\end{eqnarray}
where $\otimes$ stands for convolutions in both longitudinal and
transverse momenta of partons~\cite{CL}.  The schematic picture of the
factorization theorem is given in Fig.~\ref{fig:fact}. 
The distribution amplitudes, which are universal in the processes under
consideration, are determined from experiments, the light-cone QCD sum
rules, lattice calculations, or other theoretical methods.  The hard
kernel is characterized by a hard scale $Q\sim\sqrt{\bar\Lambda m_b}$,
where $\bar\Lambda$ is a hadronic scale and $m_b$ the $b$ quark
mass~\cite{Kurimoto:2001zj,Chen:2001pr,Mishima:2001ms}, and can be 
evaluated as an expansion in powers of $\alpha_s(Q)$ and
$\bar\Lambda/Q$.  The hard kernels of the spectator and annihilation
contributions, as well as the emission contribution, are calculable and
start from $O(\alpha_s)$.  PQCD predicts a large direct CP asymmetry in
$B^0\to\pi^\mp K^\pm$ as a result of a large strong phase arising from
annihilation penguin diagrams~\cite{KLS}.

\section{\boldmath{$B\to\pi K$} and \boldmath{$\pi\pi$}
 Puzzles\label{sec:pik}} 

The current data of the direct CP asymmetries of $B\to \pi K$ and the
branching ratios of $B\to \pi\pi$~\cite{HFAG}, 
\begin{eqnarray}
A_{CP}(B^\pm\to \pi^0 K^\pm)&=&(4\pm 4)\%
\;,\nonumber\\
A_{CP}(B^0\to \pi^\mp K^\pm) &=& (-10.8\pm 1.7)\%
\;,\nonumber\\
{\rm Br}(B^0\to\pi^\pm\pi^\mp) &=& (4.9\pm 0.4)\times 10^{-6}
\;,\nonumber\\
{\rm Br}(B^0\to\pi^0\pi^0) &=& (1.45\pm 0.29)\times 10^{-6}
\;,
\label{eq:data}
\end{eqnarray}
have been inconsistent with the expected relations 
\begin{eqnarray}
A_{CP}(B^\pm\to \pi^0 K^\pm) &\approx& A_{CP}(B^0\to \pi^\mp K^\pm)
\;,\nonumber\\
{\rm Br}(B^0\to\pi^\pm\pi^\mp) &\gg& {\rm Br}(B^0\to\pi^0\pi^0)
\;.
\label{eq:rel1}
\end{eqnarray} 
These relations can be understood in the topological-amplitude
decompositions~\cite{Chau:1990ay}: 
\begin{eqnarray}
{\cal A}(B^+\to \pi^+ K^0) &=&
P'
\nonumber\\
\sqrt{2}{\cal A}(B^+\to \pi^0 K^+) &=& 
-P'-P'_{ew}
-\left(T'+C' \right)e^{i\phi_3}
\;,\nonumber\\
{\cal A}(B^0\to \pi^- K^+) &=& 
-P'-T'e^{i\phi_3}
\;,\nonumber\\
\sqrt{2}{\cal A}(B^0\to \pi^0K^0) &=&
P' - P'_{ew}
-C'e^{i\phi_3}
\;,
\end{eqnarray}
and 
\begin{eqnarray}
{\cal A}(B^0\to \pi^+\pi^-) &=&
-T-Pe^{i\phi_2}
\;,\nonumber\\
\sqrt{2}{\cal A}(B^+\to \pi^+\pi^0) &=& 
-T - C
-P_{ew}e^{i\phi_2}
\;,\nonumber\\
\sqrt{2}{\cal A}(B^0\to \pi^0\pi^0) &=& 
-C+\left(P-P_{ew}\right)e^{i\phi_2}
\;,
\end{eqnarray}
where $T^{(\prime)}$, $C^{(\prime)}$, $P^{(\prime)}$, and
$P_{ew}^{(\prime)}$ stand for the color-allowed tree, color-suppressed
tree, penguin, and electroweak penguin amplitudes, respectively, and
$\phi_2$ and $\phi_3$ are the weak phase defined by
$V_{ub}=|V_{ub}|\exp(-i\phi_3)$, $V_{td}=|V_{td}|\exp(-i\phi_1)$, and
$\phi_2=180^\circ - \phi_1 - \phi_3$.  Assuming the hierarchies,  
$P' > T',\ P'_{ew} > C'$ for $B\to\pi K$ and $T > C,\ P > P_{ew}$ for 
$B\to\pi\pi$~\cite{Yoshikawa,GHLR,Charng:2003iy}, the relations in 
Eq.~(\ref{eq:rel1}) can be derived.  The current data seem to require a
large electroweak penguin amplitude~\cite{Gronau:2003kj,Yoshikawa, 
BFRS,Barger:2004hn,Nandi:2004dx,FOURTH,Wu:2005hi},
a large color-suppressed tree
amplitude~\cite{Chiang:2004nm,Ligeti:2004ak,Charng:2004ed,He:2004ck}, or
both of them~\cite{Wu:2004xx,Baek:2004rp,Kim:2005jp}.

The LO PQCD predictions follow the na\"{\i}ve expectations as shown in
Tables~\ref{tab:br} and \ref{tab:acp}~\cite{KLS,Lu:2000em,Keum:2003qi}. 
\begin{table}[h]
\begin{center}
\caption{Branching ratios for the $B\to \pi K$ and $\pi\pi$ decays in
 units of $10^{-6}$.  
\label{tab:br}}
\begin{tabular}{|c|c|c|c|}
\hline 
\textbf{Mode} & \textbf{Data}~\cite{HFAG} & 
\textbf{LO}~\cite{Keum:2003qi} & \textbf{NLO}~\cite{Li:2005kt,Li:2006cv}
\\
\hline
$B^\pm\to \pi^\pm K^0$ &
 $24.1 \pm 1.3$ &
 $14.4 \sim 26.3$ &
 $23.6^{+14.5}_{-\ 8.4}$
\\
$B^\pm\to\pi^0 K^\pm$ &
 $12.1 \pm 0.8$ &
 $7.9 \sim 14.2$ &
 $13.6^{+10.3}_{-\ 5.7}$
\\
$B^0\to\pi^\mp K^\pm$ & 
 $18.9 \pm 0.7$ &
 $12.7 \sim 19.3$ &
 $20.4^{+16.1}_{-\ 8.4}$
\\
$B^0\to \pi^0 K^0$ & 
 $11.5 \pm 1.0$ &
 $4.5 \sim 8.1$ &
 $8.7^{+\ 6.0}_{-\ 3.4}$
\\
\hline
$B^0\to\pi^\mp\pi^\pm$ & 
 $4.9 \pm 0.4$ &
 $5.9 \sim 11.0$ &
 $6.5^{+\ 6.7}_{-\ 3.8}$
\\
$B^\pm\to\pi^\pm\pi^0$ &
 $5.5 \pm 0.6$ &
 $2.7 \sim 4.8$ &
 $4.0^{+\ 3.4}_{-\ 1.9}$
\\
$B^0\to\pi^0\pi^0$ &
 $1.45 \pm 0.29$ &
 $0.10 \sim 0.65$ &
 $0.29^{+0.50}_{-0.20}$
\\
\hline
\end{tabular}
\end{center}
\end{table}
\begin{table}[h]
\begin{center}
\caption{Direct CP asymmetries for the $B\to \pi K$ and $\pi\pi$ decays
 in percentage. 
\label{tab:acp}}
\begin{tabular}{|c|c|c|c|}
\hline 
\textbf{Mode} & \textbf{Data}~\cite{HFAG} & 
\textbf{LO}~\cite{Keum:2003qi} & \textbf{NLO}~\cite{Li:2005kt,Li:2006cv}
\\
\hline
$B^\pm\to \pi^\pm K^0$ &
 $-2 \pm 4$ & 
 $-1.5 \sim -0.6$ &
 $0 \pm 0$
\\
$B^\pm\to\pi^0 K^\pm$ &
 $4 \pm 4$ & 
 $-17.3 \sim -10.0$ &
 $-1^{+3}_{-6}$
\\
$B^0\to\pi^\mp K^\pm$ & 
 $-10.8 \pm 1.7$ & 
 $-21.9 \sim -12.9$ &
 $-10^{+7}_{-8}$
\\
$B^0\to \pi^0 K^0$ & 
 $2 \pm 13$ &
 $-1.03 \sim -0.90$ &
 $-7^{+3}_{-4}$
\\
\hline
$B^0\to\pi^\mp\pi^\pm$ &
 $37 \pm 10$ & 
 $16.0 \sim 30.0$  &
 $18^{+20}_{-12}$
\\
$B^\pm\to\pi^\pm\pi^0$ &
 $1 \pm 6$&
 $0.0 $ &
 $0 \pm 0$
\\
$B^0\to\pi^0\pi^0$ &
 $28^{+40}_{-39}$ &
 $20.0 \sim 40.0$ &
 $63^{+35}_{-34}$
\\
\hline
\end{tabular}
\end{center}
\end{table}
Because LO PQCD predicts a negligible $C^{(\prime)}$, NLO contributions 
may enhance $C^{(\prime)}$ sufficiently.  Recently, the most important
NLO contributions from the vertex corrections, the quark loops, and the
magnetic penguin were calculated in the PQCD
approach~\cite{Li:2005kt,Li:2006cv}.  The sum of the latter two reduces
the LO penguin amplitudes by about 10\% in the $B\to\pi K$ decays and
affects the CP asymmetries little.  The vertex corrections enhance $C'$
by a factor of three and rotate the phase of $C'$, so that $T'+C'$ is
parallel to $P'+P'_{ew}$.  Thus, the direct CP asymmetry of
$B^\pm\to\pi^0 K^\pm$ vanishes as shown in Table~\ref{tab:acp}.

\begin{figure*}[t]
\centering
\includegraphics{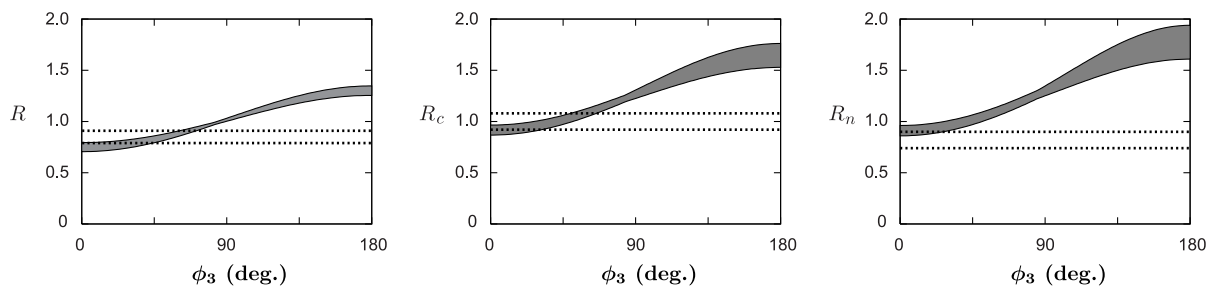}
\caption{$R$, $R_c$, and $R_n$ as functions of $\phi_3$ from NLO PQCD
 with the bands representing the theoretical
 uncertainty~\cite{Li:2005kt}.  The two dashed lines denote $1\sigma$ 
 bounds from the data. 
\label{fig:ratios}}
\end{figure*}

To compare the predicted $B\to\pi K$ branching ratios with the data, it
is useful to consider the ratios 
\begin{eqnarray}
R &=&
\frac{{\rm Br}(B^0\to\pi^\mp K^\pm)}{{\rm Br}(B^\pm\to\pi^\pm K^0)}
\frac{\tau_{B^+}}{\tau_{B^0}}
\ =\ 0.85\pm 0.06
\;,\nonumber\\
R_c &=&
2\, \frac{{\rm Br}(B^\pm\to\pi^0 K^\pm)}{{\rm Br}(B^\pm\to\pi^\pm K^0)}
\ =\ 1.00\pm 0.08
\;,\nonumber\\
R_n &=&
\frac{1}{2}\frac{{\rm Br}(B^0\to\pi^\mp K^\pm)}{{\rm Br}(B^0\to\pi^0 K^0)}
\ =\ 0.82\pm 0.08
\;.
\end{eqnarray}
As shown in Fig.~\ref{fig:ratios}, the PQCD predictions for $R$ and
$R_c$ are consistent with the data, whereas that for $R_n$ is larger
than the data~\cite{Li:2005kt}. 
The larger $R_n$ is due to the smaller $B^0\to\pi^0 K^0$ branching
ratio, which might imply a significant $P'_{ew}$.  However, the
discrepancy is not serious at this moment, because the theoretical
uncertainty is representative and the actual uncertainty could be
larger.

The enhanced $C'$ also affects the mixing-induced CP asymmetry of
$B^0\to\pi^0 K_S$, given by 
\begin{eqnarray}
S_{\pi^0K_S} \ =\
{2\,\rm{Im}(\lambda_{\pi^0K_S}) \over 1+|\lambda_{\pi^0K_S}|^2}
\;,
\end{eqnarray}
with 
\begin{eqnarray}
\lambda_{\pi^0 K_S} \ =\
-e^{-2i\phi_1} 
{P' -P'_{ew} - C'e^{-i\phi_3} \over P' -P'_{ew} - C'e^{i\phi_3}}
\;.
\end{eqnarray}
$S_{\pi^0K_S}$ may deviate from $S_{c\bar cs}=\sin(2\phi_1)$ in 
$b\to c\bar c s$ modes.  Therefore, it is essential to include the NLO
corrections in the calculation of $S_{\pi^0 K_S}$.  It was found that
the predicted possible deviation of $S_{\pi^0 K_S}$ is still small, 
$S_{\pi^0 K_S} - S_{c\bar c s} = 0.06^{+0.02}_{-0.03}$~\cite{Li:2005kt}.
Moreover, PQCD predicts a positive deviation, opposite to the observed
one.

$C$ for $B\to\pi\pi$, as well as $C'$ for $B\to\pi K$, is enhanced by
the vertex corrections, but it is insufficient to accommodate the
$B^0\to\pi^0\pi^0$ branching ratio to the measured
value~\cite{Li:2005kt}.  NLO PQCD predicts $|C/T|\approx 0.2$ for
$B\to\pi\pi$, though a much larger $|C/T|\approx 0.8$ is required to
explain the observed $B^0\to\pi^0\pi^0$ branching
ratio~\cite{Charng:2004ed}.

The same NLO PQCD formalism was applied to the $B\to\rho\rho$
decays~\cite{Li:2006cv}, which are sensitive to the color-suppressed
tree contribution.   The predicted $B\to\rho\rho$ branching ratios are
listed in Table~\ref{tab:br2}. 
\begin{table}[h]
\begin{center}
\caption{Branching ratios for the $B\to\rho\rho$ decays in units of
 $10^{-6}$~\cite{Li:2006cv}. 
\label{tab:br2}}
\begin{tabular}{|c|c|c|c|c|}
\hline 
\textbf{Mode} & \textbf{BABAR}~\cite{HFAG} & \textbf{Belle}~\cite{HFAG}
& \textbf{LO} & \textbf{NLO}
\\
\hline
$B^0\to\rho^\mp \rho^\pm$ & 
 $30 \pm 4\pm 5$ &
 $22.8 \pm 3.8^{+2.3}_{-2.6}$ &
 $27.8$&
 $25.3^{+25.3}_{-13.8}$
\\
$B^\pm\to\rho^\pm \rho^0$ & 
 $17.2 \pm 2.5 \pm 2.8$ &
 $31.7 \pm 7.1^{+3.8}_{-6.7}$ &
 $13.7$&
 $16.0^{+15.0}_{-\ 8.1}$
\\
$B^0\to\rho^0 \rho^0$ & 
 $ < 1.1$ &
 --- &
 $0.33$&
 $0.92^{+1.10}_{-0.56}$
\\
\hline
\end{tabular}
\end{center}
\end{table}
The LO results differ from those in the previous LO
analyses~\cite{Li:2005hg,Chen:2006jz} slightly due to the different
choices of the hard scale and parameters.  The NLO PQCD predictions for
the $B^0\to\rho^\mp\rho^\pm$ and $B^\pm\to\rho^\pm\rho^0$ branching
ratios are consistent with the data.  Because the decay amplitudes for
$B\to\rho\rho$ are similar to those for $B\to\pi\pi$, the branching 
ratio of $B^0\to\rho^0\rho^0$ is expected to be larger than that of
$B^0\to\pi^0\pi^0$ due to the meson decay constants $f_\rho > f_\pi$.
In fact, the NLO predictions follow this expectation, and the central
value of the predicted $B^0\to\rho^0\rho^0$ branching ratio has almost
reached the experimental upper bound.  The NLO PQCD analysis has thus
confirmed that it is unlikely to accommodate both the $B^0\to\pi^0\pi^0$
and $B^0\to\rho^0\rho^0$ branching ratios to the data simultaneously.
Hence, the $B\to\pi\pi$ puzzle is confirmed in the PQCD approach.  All
proposed resolutions to the $B\to\pi\pi$ puzzle should survive the
constraints from the $B^0\to\rho^0\rho^0$ data.

I comment on $B\to\pi K$ and $\pi\pi$ results in other theoretical
approaches.  In QCDF, $C^{(\prime)}$ is enhanced by the NLO jet function
obtained from SCET and the large $B^0\to\pi^0\pi^0$ branching ratio can
be explained~\cite{Beneke:2005gs}.  However, the inclusion of the NLO
jet function overshoots the $B^0\to\rho^0\rho^0$ branching ratio and
deteriorates the predictions for the $B^\pm\to\pi^0 K^\pm$ and
$B^0\to\pi^\mp K^\pm$ direct CP asymmetries~\cite{Li:2006cv}.  In SCET,
incalculable soft contributions are regarded as phenomenological
parameters, which can be fitted from the experimental data.  It was
found that the charming penguin, which is one of the phenomenological
parameters, is large as $|C^{(\prime)}/T^{(\prime)}|\approx
1$~\cite{Bauer:2004tj,Bauer:2005kd,Williamson:2006hb}.  Consequently,
the large $B^0\to\pi^0\pi^0$ branching ratio can be realized.  The
$B^0\to\rho^0\rho^0$ branching ratio should be checked in the same
formalism.  The ratio $C^{(\prime)}/T^{(\prime)}$ is, however, real in
the leading-power SCET formalism, and therefore the $B^\pm\to\pi^0
K^\pm$ and $B^0\to\pi^\mp K^\pm$ direct CP asymmetries can not be
explained at the same
time~\cite{Bauer:2004tj,Bauer:2005kd,Williamson:2006hb}.

\section{Polarizations in \boldmath{$B\to VV$}
\label{sec:vv}}

In the na\"{\i}ve factorization approximation, the longitudinal and
transverse polarization fractions, $R_L$ and $R_{\parallel,\perp}$,
respectively, in $B\to VV$ modes obey the power-counting
rules~\cite{AK},  
\begin{eqnarray}
R_L\, \sim\, 1- O(m_V^2/m_B^2)
\;, \ \ 
R_{\parallel,\perp}\, \sim\, O(m_V^2/m_B^2)
\;, 
\label{eq:VV}
\end{eqnarray}
where $m_B$ is the mass of the $B$ meson and $m_V$ that of the emitted
vector meson from the weak vertex.  The data of the longitudinal
polarization fractions are given by~\cite{HFAG} 
\begin{eqnarray}
R_L(B^0\to \rho^+\rho^-) &=& 0.967^{+0.023}_{-0.028}
\;,\nonumber\\
R_L(B^+\to \rho^+\rho^0) &=& 0.96 \pm 0.06
\;,\nonumber\\
R_L(B^+\to K^{*+}\rho^0) &=& 0.91^{+0.23}_{-0.21}
\;,\nonumber\\
R_L(B^+\to K^{*0}\rho^+) &=& 0.48^{+0.09}_{-0.08}
\;,\nonumber\\
R_L(B^+\to \phi K^{*+}) &=& 0.50 \pm 0.07 
\;,\nonumber\\
R_L(B^0\to \phi K^{*0}) &=& 0.48 \pm 0.04 
\;.
\end{eqnarray}
The polarization fractions for tree-dominated modes, $B\to\rho\rho$,
satisfy Eq.~(\ref{eq:VV}).  However, those for penguin-dominated modes,
$B\to\phi K^*$ and $K^{*0}\rho^+$, obviously conflict with the na\"{\i}ve
expectations.

For the penguin-dominated modes, the polarization fractions could be
modified by sub-leading contributions.  The penguin annihilation
contribution from the $(S-P)(S+P)$ operators, which follows 
$R_L\sim R_\parallel\sim R_\perp$, could decrease the longitudinal
fraction~\cite{AK}.  In LO PQCD, the spectator and penguin annihilation
contributions help to reduce $R_L$ as shown in Table~\ref{tab:pol}, but
it is not enough to explain the $B\to\phi K^*$ and $K^{*0}\rho^+$
data~\cite{Chen:2002pz,Li:2004ti,Huang:2005if}. 
\begin{table}[h]
\begin{center}
\caption{Polarization fractions in $B\to\phi K^*$, (I) without spectator
 and annihilation contributions, and (II) with spectator and
 annihilation contributions~\cite{Chen:2002pz}.  
\label{tab:pol}}
\begin{tabular}{|c|c|c|c|c|c|c|}
\hline 
\textbf{Mode} & & \boldmath{$R_{L}$} & \boldmath{$R_{\parallel}$} &
 \boldmath{$R_{\perp}$} & \boldmath{$\phi_{\parallel}{\rm (rad)}$} &
 \boldmath{$\phi_{\perp}{\rm (rad)}$} 
\\
\hline
$B^0\to\phi K^{*0}$ 
 & (I) &  $0.923$ & $0.040$ & $0.035$ & $\pi$  & $\pi$  \\
 & (II) & $0.750$ & $0.135$ & $0.115$ & $2.55$ & $2.54$ \\
\hline 
$B^+\to\phi K^{*+}$ 
 & (I) &  $0.923$ & $0.040$ & $0.035$ & $\pi$  & $\pi$  \\
 & (II) & $0.748$ & $0.133$ & $0.111$ & $2.55$ & $2.54$ \\
\hline
\end{tabular}
\end{center}
\end{table}
$R_L$ for $B^+\to K^{*+}\rho^0$ remains as $R_L\sim 0.85$, which is
consistent with the data, because this process involves additional
tree amplitudes.  The tree-dominated modes, which are insensitive to the
sub-leading corrections, follow the na\"{\i}ve counting rules in
Eq.~(\ref{eq:VV})~\cite{Li:2005hg,Li:2006cv}.  Therefore, only the
penguin-dominated modes $B\to\phi K^*$ and $K^{*0}\rho^+$ have exhibited
anomalies in the measured polarization fractions.

There have been several mechanisms proposed to explain the observed 
$B\to\phi K^*$ polarizations.  In PQCD, it was proposed that the 
$B\to K^*$ form factor $A_0$, associated with the longitudinal
polarization, may be smaller than the central value of the LO PQCD
prediction~\cite{Li:2004mp}.  Postulating a smaller value, 
$A_0\approx 0.3$, which does not contradict to any existing 
measurements, $R_L$ for $B\to\phi K^*$ decreases to $0.6$.
Another mechanism in PQCD was proposed in Ref.~\cite{Chen:2006jz}.
Adopting a modified definition of the hard scale, which is a source of
theoretical uncertainty, $R_L$ for $B\to\phi K^*$ could approach to
$0.6$.  However, $R_L$ for $B^+\to K^{*0}\rho^+$ becomes about $0.8$,
which is inconsistent with the data. This is because the sign of the
real part of the annihilation amplitude for $B^+\to K^{*0}\rho^+$ is 
opposite to that for $B\to\phi K^*$~\cite{Chen:2006jz}.  Small $R_L$ for
the penguin-dominated modes might come from the complicated QCD
dynamics, but it is important to explain both the $B\to\phi K^*$ and
$K^{*0}\rho^+$ data.

\section{Summary\label{sec:summary}}

In this talk, I have summarized the recent works on exclusive hadronic
$B$ meson decays in the PQCD approach, concentrating on the observed
deviations in the branching ratios and the CP asymmetries of the
$B\to\pi K$, $\pi\pi$ modes, and the polarization fractions of
penguin-dominated $B\to VV$ modes, which are sensitive to subdominant
contributions.

Including the important NLO contributions, the color-suppressed tree
amplitude is enhanced by the vertex corrections, and therefore the
predicted direct CP asymmetries of the $B\to\pi K$ modes become
consistent with the experimental data.  However, it is unlikely to
accommodate both the $B^0\to\pi^0\pi^0$ and $B^0\to\rho^0\rho^0$
branching ratios to the measured ones simultaneously.

The polarization fractions of the penguin-dominated $B\to VV$ modes
deviate from the na\"{\i}ve power-counting rules, including the spectator
and annihilation contributions.  However, it is not enough to explain
the observed data.  A small longitudinal fraction for $B\to\phi K^*$
might come from QCD uncertainty, but it is necessary to explain both 
the $B\to\phi K^*$ and $K^{*0}\rho^+$ data.  NLO corrections to the
polarization fractions should be studied in future work.

\bigskip 
\begin{acknowledgments}
I am grateful to the organizers of FPCP2006 for inviting me to the
 exiting conference.  I thank C.H.~Chen, W.S.~Hou, H-n.~Li, and
 A.I.~Sanda for fruitful and helpful discussions.  This work was
 supported by the U.S. Department of Energy under Grant
 No. DE-FG02-90ER40542. 
\end{acknowledgments}

\bigskip 

\end{document}